\documentstyle[aps]{revtex}
\def\be{\begin{equation}}
\def\ee{\end{equation}}
\newcommand{\pp}{\ \ .}

\newcommand\ket[1]{\mathop{\left| #1\right\rangle}\nolimits}
\newcommand{\dd}{\mathop{\rm d\mkern-2mu}\nolimits}
\newcommand{\AmS}{{\protect\the\textfont2
  A\kern-.1667em\lower.5ex\hbox{M}\kern-.125emS}}

\title{\Large Solutions to a Quantal Gravity-Matter Field Theory on a Line}
\author{Roman Jackiw\address{\footnotesize Center for Theoretical Physics,
Massachusetts
              Institute of Technology\\ 
              6-320, 77 Massachusetts Avenue, Cambridge MA 02139-4307,
USA}\protect\thanks{This work is supported in part by funds provided by
the U.S. Department of Energy under contract
\#DE-FC02-94ER40818. 
\quad MIT-CTP-2598}}
       
\begin{document}

\maketitle
\medskip

\centerline{\large Dedicated to Tullio Regge on his sixty-fifth birthday}
\medskip
\centerline{\large\it 2$^{\it nd}$ Conference on Constrained Dynamics and
Quantum Gravity}
\centerline{\it Santa Margherita,  Italy, September 1996}
\bigskip

\begin{abstract}\noindent{\bf Abstract.}\quad
Solutions to a scalar-tensor (dilaton) quantum gravity
theory, interacting with quantized matter, are described. Dirac quantization is frustrated
by quantal anomalies in the constraint algebra. Progress is made only
after the Wheeler--DeWitt equation is modified by quantal terms, which
eliminate the anomaly. More than one modification is possible, resulting
in more than one `physical' spectrum in the quantum theory,
corresponding to the given classical model.
\end{abstract}

\setlength{\baselineskip}{1.5\baselineskip}  %% to be removed for final
\mathsurround1pt

\section{Introduction}\markboth{Solutions to a Quantal Gravity-Matter Field Theory on a
Line}{
Introduction}%
Constructing the quantum version of classical gravity theory
remains a challenge for theoretical physics, because Einstein's theory
resists quantization for a variety of reasons. The principal obstacle is its
nonrenormalizability, and this prevents meaningful calculation of quantal
processes. Additionally, there are conceptual difficulties that are peculiar
to diffeomorphism-invariant quantum theories. These include the
question of how to introduce time into the theory, the interpretational
issues that arise from the unfamiliar role of the Hamiltonian as a
constraint, and the puzzles of unitary evolution in presence of classical
singularities (black holes). Very little illumination can be gotten on these
topics, when explicit calculation is frustrated by infinities.

One response to the impasse of quantum gravity is to look at models in dimensions lower
than the physical (3+1). In the lower-dimensional worlds there are no propagating
gravitons, and no associated infinities. But the conceptual questions surrounding a
diffeomorphism-invariant quantum theory remain. Tullio Regge, to whom this essay is
dedicated  on his sixty-fifth birthday, has actively advanced the subject of
(2+1)-dimensional gravity by exploring its mathematical, specifically its algebraic,
structures. This model has the added attraction that it is
physically relevant as a description of processes that are kinematically
restricted to move on a plane in the presence of cosmic strings. Also it gives a realization to the
Regge link-calculus for geometry.

I shall describe some research carried out with colleagues on  still lower dimensional
gravity -- gravity on a line or {\em lineal} gravity, {\it i.e.}, in (1+1) dimensions.

The approach that I take is to view gravity as a gauge theory. But when I speak
of gravity theory as a `gauge theory', I mean only that the symmetry
transformations, which leave the model invariant, depend on parameters
that are arbitrary functions of space-time, as is clearly the case with
invariance against arbitrary coordinate redefinition. I do not necessarily
imply that gravity theory can be formulated as a vectorial gauge theory. 

In fact, for lower-dimensional gravity it is frequently possible to formulate the model as
a vectorial gauge theory, based on a finite-dimensional Lie group, but with kinematics
not following the Yang--Mills paradigm; rather, other -- non-Yang--Mills --
kinetic 
terms, available in these lower dimensions, are employed: Chern--Simons term
in (2+1)~dimensions, B--F expression in (1+1)~dimensions.

However, in my presentation here I shall not delve into such specifically gauge-theoretic
formulations of gravity theory, but I shall organize my presentation around the theme
common to all gauge theories: the existence of constraints in a canonical
(symplectic) formulation, and the possible emergence of quantal obstructions --
anomalies -- in the quantum algebra of constraints. 
My hope is that the experience gained in the last quarter-century with vectorial gauge
theories will enable us to go far in the analysis of the gravitational model. 

\section{Vector Gauge Theories}\markboth{Solutions to a Quantal Gravity-Matter Field Theory
on a Line}{
Vector Gauge Theories}%
In a canonical, Hamiltonian approach to quantizing a
theory with local symmetry -- a theory that is invariant against
transformations whose parameters are arbitrary functions on space-time
--  there occur constraints, which are imposed on physical states.  
Typically these constraints correspond to time components of the
Euler--Lagrange equations, and 
vectorial gauge theories provide
familiar examples.   The time
component of a Yang--Mills gauge field equation is the Gauss law:
\be
G_a \equiv {\bf D} \cdot {\bf E}^a - \rho^a = 0\pp
\ee
Here ${\bf E}^a$ is the (non-Abelian) electric field, $\rho^a$ the
matter charge density, and ${\bf D}$ denotes the gauge-covariant
derivative.  When expressed in terms of canonical variables, $G_a$ does
{\it not\/} involve time-derivatives -- it depends on canonical
coordinates and momenta, which we denote collectively by the symbols $X$
and $P$, respectively  ($X$ and $P$ are fields defined at fixed time):
$G_a = G_a (X,P)$.   Thus in a Schr\"odinger representation for the
theory, the Gauss law condition on physical states
\be
G_a (X,P) | \, \psi \, \rangle = 0
\ee
corresponds to a (functional)  differential  
 equation that the state
functional $\Psi(X)$ must satisfy:
\be
G_a \left( X, {1\over i} {\delta \over \delta X} \right)
\, \Psi(X) = 0\pp
\label{eq:3}
\ee
In fact, Eq.\,(\ref{eq:3}) 
 represents an infinite number of equations,
one for each spatial point ${\bf r}$, since $G_a$ is also the generator
of the local symmetry:
$G_a = G_a({\bf r})$.
Consequently, questions of consistency (integrability) arise, and these
may be examined by considering the commutator of two constraints.
Precisely because the $G_a$ generate the symmetry transformation,
one expects their commutator to follow the Lie algebra with structure
constants $f_{abc}$:
\be
\left[ G_a ({\bf r}) , G_b (\tilde{\bf r}) \right]
= i \, f_{abc} \, G_c ({\bf r}) \, \delta({\bf r} - \tilde{\bf r}) \pp
\label{eq:4}
\ee
If (\ref{eq:4}) holds, the constraints are consistent -- they are first-class -- and the
constraint equations (\ref{eq:3}) are integrable, at least locally.

However, it is by now well known that Eq.\,(\ref{eq:4}), which {\it
does\/} hold classically with Poisson bracketing, may acquire a quantal
anomaly.  Indeed, when the matter charge density is constructed from
fermions of a definite chirality, the Gauss law algebra is modified by
an extension -- a Schwinger term -- the constraint equations become
second-class and Eq.\,(\ref{eq:3}) is inconsistent and cannot be solved.
We call such gauge theories `anomalous'.

This does not mean that a quantum theory cannot be constructed from an
anomalous gauge theory.  One can adopt various strategies for overcoming
the obstruction, but these represent modifications of the original
model.  Moreover, the resulting quantum theory possesses physical
content that 
can be
 far removed from what one might infer by studying
the classical model.  All this is explicitly illustrated by the
anomalous chiral Schwinger model, whose Gauss law is obstructed, while a
successful construction of the quantum theory leads to massive
excitations, which cannot be anticipated from the unquantized equations
[1].

\section{Gravity Theories}\markboth{Solutions to a Quantal Gravity-Matter Field Theory on a
Line}{
Gravity Theories}%
With these facts in mind, we turn now to gravity theory, which obviously
is invariant against local transformations that redefine coordinates of
space-time.
It is found that here too the
 time components of Einstein's equation
again
 comprise
the constraints
in a canonical formulation of dynamics:
\be
{1\over   8\pi G}  \left( R_\nu^{\,0} - 
{\textstyle{1\over2}} \delta_\nu^{\,0} R \right)
- T_\nu^{\,0} = 0\pp
\label{eq:5}
\ee
The gravitational  part is the time component of the Einstein tensor 
$R^\mu_{\nu} - {1\over2} \delta^\mu_{\nu} R$;
weighted by Newton's constant $G$, this equals the time component of the
matter energy-momentum tensor, $T^{\mu}_{\nu}$.   In the quantized
theory, the collection of canonical operators on the left side in
(\ref{eq:5}) 
should
annihilate physical states.  The resulting equations may be
presented as
\begin{eqnarray}
   {\cal E} \ket\psi  &=& 0
   \label{eq:6} \\
   {\cal P}_i \ket\psi  &=& 0 
   \label{eq:7}
\end{eqnarray}
where ${\cal E}$ is the energy constraint
[the $\nu=0$ component of (\ref{eq:5})], composed of its gravity and
matter parts
\be
  {\cal E} = {\cal E}^{\rm\,gravity} + {\cal E}^{\rm\,matter}
\ee
and ${\cal P}_i$ is the momentum constraint 
[the $\nu=i$ component of (\ref{eq:5})], also with gravity and matter
parts:
\be
{\cal P}_i = {\cal P}_i^{\rm\,gravity} + {\cal P}_i^{\rm\,matter}\pp
\ee
Taking for definiteness matter to be described by a massless, spinless
field $\varphi$, with canonical momentum $\Pi$, we have
\begin{eqnarray}
{\cal E}^{\rm\,matter} &=& 
{\textstyle{1\over2}} \left( \Pi^2 + \gamma \, 
\gamma^{ij} \, \partial_i  \varphi \, \partial_j \varphi \right) \\
{\cal P}_i^{\rm\,matter} &=& \partial_i  \varphi \, \Pi \pp
\end{eqnarray}
Here $\gamma_{ij}$ is the spatial metric tensor;
$\gamma$  its determinant;
$\gamma^{ij}$  its inverse.

The momentum constraint in Eq.\,(\ref{eq:7}) is easy to unravel.
In a Schr\"odinger representation, it requires that 
$\Psi(\gamma_{ij}, \varphi)$ 
be a functional of the canonical field variables
$\gamma_{ij},\varphi$ 
that is invariant against reparameterization of the spatial
coordinates, and such functionals are easy to construct.

Of course it is (\ref{eq:6}), the Wheeler--DeWitt equation, that is
highly nontrivial, and once again one asks about its consistency.  If
the commutators of ${\cal E}$ and ${\cal P}$ follow their Poisson
brackets one would expect
that the following algebra holds:
 %%\begin{mathletters}%
 %% \label{eq:12all}
\begin{eqnarray}%
\left[ {\cal P}_i ({\bf r}), {\cal P}_j (\tilde{\bf r}) \right]
&=& i {\cal P}_j ({\bf r}) \, \partial_i \, 
\delta({\bf r} - \tilde{\bf r}) %%\nonumber\\
  %% &&{}
+ i {\cal P}_i (\tilde{\bf r}) \, 
\partial_j \, \delta({\bf r} - \tilde{\bf r})
\label{eq:12a} \\
\left[ {\cal E} ({\bf r}), {\cal E} (\tilde{\bf r}) \right]
&=& i \Bigl( {\cal P}^i ({\bf r}) + 
{\cal P}^i (\tilde{\bf r}) \Bigr) \partial_i \, \delta({\bf r} - \tilde{\bf
r})      %\nonumber\\ &&{}
\label{eq:12b} \\
\left[ {\cal E} ({\bf r}), {\cal P}_i (\tilde{\bf r}) \right]
&=& i \Bigl( {\cal E} ({\bf r}) + 
{\cal E} (\tilde{\bf r}) \Bigr) \partial_i \, \delta({\bf r} - \tilde{\bf r})
\pp      %\nonumber\\ &&{}
\label{eq:12c}
\end{eqnarray}%
 %%\end{mathletters}%
Here ${\cal P}^i \equiv \gamma \, \gamma^{ij} \, {\cal P}_j$.
If true, Eqs.\,(\ref{eq:12a})--(\ref{eq:12c}) would demonstrate the consistency of the
constraints, since they appear first-class.
In fact, since we have already described the solution to the momentum
constraint (\ref{eq:7}), we expect that (\ref{eq:12a}) is not modified by
quantal corrections. It remains to examine (\ref{eq:12b})--(\ref{eq:12c}).
  Unfortunately, establishing 
(\ref{eq:12b})--(\ref{eq:12c}) in the quantized theory is highly
problematical.  First of all there is the issue of operator ordering in the
gravitational portion of ${\cal E}$ and ${\cal P}$.  Much has been said
about this, and I shall not address that difficulty here. 

The problem that I want to call attention to is the very likely
occurrence of an extension in the $[{\cal E}, {\cal P}_i]$ commutator
(\ref{eq:12c}).  We know that in flat space, the commutator between the
matter energy and momentum densities  possesses a noncanonical triple
derivative Schwinger term\,\cite{2}.  There does not appear any known mechanism
arising from the gravity variables
that would effect a cancellation of this obstruction.

A definite resolution of this question in the full quantum theory is out
of reach at the present time.  Non-canonical Schwinger terms can be
determined only after a clear understanding of the singularities in the
quantum field theory and the nature of its Hilbert space are in hand,
and this is obviously lacking for four-dimensional quantum gravity.

Faced with the impasse, we turn to a gravitational model in
two-dimensional space-time
-- a {\it lineal\/} gravity theory --
where the calculation can be carried to a
definite conclusion: an obstruction does exist and the model is
anomalous.  Various mechanisms are available to overcome the anomaly,
but the resulting various quantum theories 
are inequivalent.

In two dimensions, Einstein's equation is vacuous because 
$R_\nu^\mu = {1\over2} \delta^\mu_\nu R$;
therefore, gravitational dynamics has to be invented afresh.   The models
that have been studied recently posit local dynamics for the `gravity'
sector, which involves as variables the metric tensor and an additional
world scalar (`dilaton' or Lagrange multiplier)  field.  Such
`scalar-tensor' theories, introduced a decade ago \cite{3}, are obtained by
dimensional reduction from higher-dimensional Einstein theory \cite{3,4}.
They should be contrasted with models where quantum fluctuations of
matter variables induce gravitational dynamics \cite{5}, which therefore are
nonlocal and do not appear to offer any insight into the questions
posed by the physical, four-dimensional theory.

The model we study is the so-called `string-inspired dilaton gravity'
-- CGHS theory \cite{6}.   The gravitational action involves the metric
tensor $g_{\mu\nu}$, the dilaton field $\phi$, and a cosmological
constant $\lambda$.  The matter action describes the coupling of a
massless, spinless field $\varphi$:
\begin{eqnarray}
I_{\rm\,gravity} &=& \int \dd^{\,2}\! x \, \sqrt{-g} \,
e^{-2\phi} %%\nonumber\\ 
  %% &&{}\times
\left( R + 4 g^{\mu\nu} \partial_\mu \phi\, \partial_\nu \phi - \lambda
\right) \label{eq:13} \\
I_{\rm\,matter} &=& {\textstyle{1\over2}} \int \dd^{\,2}\! x \, 
\sqrt{-g} \, g^{\mu\nu} \, \partial_\mu \varphi\, \partial_\nu \varphi \pp
\label{eq:14}
\end{eqnarray}
The total action is the sum of (\ref{eq:13}) and (\ref{eq:14}),
weighted by `Newton's' constant $G$:
\be
I = {1\over 4\pi G} ~ I_{\rm\,gravity} + I_{\rm\,matter}\pp
\label{eq:15}
\ee

In fact this theory can be given a gauge-theoretical `B--F'
description
based on the 
centrally extended 
Poincar\'e group 
in (1+1) dimensions \cite{7}.
This formulation aided us immeasurably    in the subsequent
analysis/transformations.
However, I shall not discuss this here,
because in retrospect it proved possible
to carry the analysis forward within the 
metric formulation (\ref{eq:13})--(\ref{eq:15}).

After a remarkable sequence of redefinitions and canonical
transformations on the dynamical variables in (\ref{eq:13})--(\ref{eq:15}), 
one can present $I$ in terms
of a first-order Lagrange density ${\cal L}$ that is a sum of
 terms, quadratic in the dynamical variables\,\cite{7}:
\begin{eqnarray}
{\cal L} &=& \pi_a \dot{r}^a + \Pi \dot{\varphi} 
- \alpha {\cal E} - \beta {\cal P} 
\label{eq:16} \\
{\cal E} &=& - {\textstyle{1\over2}}
\left( 
{\textstyle{1\over \Lambda}} \pi^a \pi_a + \Lambda {r^a}' {r_a}' \right) + {\textstyle{1\over2}} \left( \Pi^2 + {\varphi'}^2 \right) 
\label{eq:17} \\
{\cal P} &=& - {r^a}' \pi_a  - \varphi' \Pi \pp
\label{eq:18}
\end{eqnarray}
I shall not derive this, but merely explain it.
The index $a$ runs over flat two-dimensional $(t,\sigma)$ space, with
signature $(1,-1)$.  Dots (dashes) signify differentiation with respect to
time $t$ 
 (space~$\sigma$).  The four variables
$\left\{ r^a, \alpha, \beta \right\}$ correspond to the four
gravitational variables $(g_{\mu\nu}, \phi)$, where only $r^a$ is
dynamical with canonically  conjugate momentum $\pi_a$, while $\alpha$
and $\beta$ act as Lagrange multipliers.  
Notice that regardless of the
sign $\Lambda \equiv \lambda / 8 \pi G$, the gravitational
contribution %%%\vadjust{\newpage}
 to ${\cal E}$, is quadratic with indefinite sign: 
 %%\begin{mathletters}%
\begin{eqnarray}
{\cal E}^{\rm\,gravity} &=&
- {\textstyle{1\over2}} \left( 
{\textstyle{1\over\Lambda}} \pi^a \pi_a +
\Lambda {r^a}' {r_a}' \right)  \nonumber \\
&=& - 
{\textstyle{1\over2\Lambda}} (\pi_0)^2 
+ {\textstyle{1\over2\Lambda}} (\pi_1)^2 %%\nonumber \\
  %% &&\quad{}
-{\textstyle{1\over2}} \Lambda ({r^0}')^2 
+ {\textstyle{1\over2}} \Lambda ({r^1}')^2  \nonumber\\
&=& - {\cal E}_0 + {\cal E}_1\label{eq:19a} \\
{\cal E}_0 &=& {\textstyle{1\over2}} \left( 
{\textstyle{1\over\Lambda}}
(\pi_0)^{2} + \Lambda ({r^0}')^2 \right)\label{eq:19b} \\
{\cal E}_1 &=& {\textstyle{1\over 2}}
\left( {\textstyle{1\over\Lambda}}
(\pi_1)^{2} + \Lambda ({r^1}')^2 \right)\pp\label{eq:19c}
\end{eqnarray}
 %%\end{mathletters}%
 On the other hand, the gravitational contribution to the momentum
does not show alteration of sign:
 %%\begin{mathletters}%
\begin{eqnarray}
{\cal P}^{\rm\,gravity} &=& - {r^a}' \pi_a \nonumber\\
                        &=& - {r^0}' \pi_0  - {r^1}' \pi_1 \nonumber\\
                        &=& {\cal P}_0 + {\cal P}_1\label{eq:20a} \\
{\cal P}_0 &=& - {r^0}' \pi_0 \label{eq:20b}\\
{\cal P}_1 &=& - {r^1}' \pi_1 \pp\label{eq:20c}
\end{eqnarray}
 %%\end{mathletters}%

All our results, and the variety of ways that they can be obtained, are a
consequence of the sign variation between the `$0$' and `$1$' contributions
to ${\cal E}^{\rm gravity}$. 
One may understand the relative negative sign  between the two
gravitational    contributors $(a=0,1)$ as follows.   Pure   metric
gravity   in two space-time dimensions is described by three functions
collected   in $g_{\mu\nu}$.  Diffeomorphism invariance involves 2
functions, which reduce the number of variables by $2\times2$, 
{\it i.e.}, pure gravity has $3-4=-1$ degrees of freedom.  Adding the
dilaton $\phi$ gives a net number of $-1+1=0$, as in our final
gravitational Lagrangian.

The matter contribution    is the conventional expression for massless
and spinless fields:
\begin{eqnarray}
{\cal E}^{\rm\,matter} &=& {\textstyle{1\over2}} (\Pi^2 + {\varphi'}^2) \label{eq:21}\\
{\cal P}^{\rm\,matter} &=& - \varphi' \, \Pi \pp \label{eq:22}
\end{eqnarray}
With the formulation in Eqs.\,(\ref{eq:16})--(\ref{eq:22})
 of the theory (\ref{eq:13})--(\ref{eq:15}) 
 we embark upon the various quantization procedures.

The transformed theory appears very simple:  there are three independent
dynamical fields 
$\left\{ r^a, \varphi \right\}$
and together with the canonical momenta 
$\left\{ \pi_a, \Pi \right\}$
they are governed by a quadratic Ham\-iltonian, which has
no interaction terms among the three.  Similarly, the momentum comprises
noninteracting terms.  However, there remains a subtle `correlation
interaction' as a consequence of the constraint that ${\cal E}$ and
${\cal P}$ annihilate physical states, as follows from varying the
Lagrange multipliers $\alpha$ and $\beta$ in (\ref{eq:16})
\begin{eqnarray}
{\cal E} \ket\psi   &=& 0  \label{eq:23} \\
{\cal P} \ket\psi   &=& 0 \pp \label{eq:24}
\end{eqnarray}
Thus, even though ${\cal E}$ and ${\cal P}$ each are sums of
noninteracting variables, the physical states 
$\ket\psi$ are {\em not} direct products of states for the
separate degrees of freedom.  Note that Eqs.\,(\ref{eq:23}) and (\ref{eq:24})
comprise the entire physical content of the theory.   There is no need
for any further `gauge-fixing' or `ghost' variables -- this is the
advantage of the Hamiltonian formalism.

As in four dimensions, the momentum constraint  %%%-\linebreak
(\ref{eq:24})
enforces invariance
of the state functional $\Psi(r^a, \varphi)$ against spatial coordinate
transformations, while the energy constraint (\ref{eq:23})
-- the Wheeler--DeWitt
equation in the present lineal gravity context -- is highly nontrivial.

Once again one looks to the algebra 
of the constraints
to check consistency.  The reduction
of (\ref{eq:12a})--(\ref{eq:12c}) to one spatial dimension leaves (after the identification
${\cal P}_i \to - {\cal P}, \gamma \gamma^{ij}
 \to 1$)
 %%\begin{mathletters}%
\begin{eqnarray}
i [ {\cal P}(\sigma), {\cal P}(\tilde{\sigma}) ] &=&
\bigl( {\cal P}(\sigma) + {\cal P}(\tilde{\sigma}) \bigr)
\, \delta'(\sigma - \tilde{\sigma}) \label{eq:31orig}\\
i [ {\cal E}(\sigma), {\cal E}(\tilde{\sigma}) ] &=&
\bigl( {\cal P}(\sigma) + {\cal P}(\tilde{\sigma}) \bigr)
\, \delta'(\sigma - \tilde{\sigma}) \\
i [ {\cal E}(\sigma), {\cal P}(\tilde{\sigma}) ] &=&
\bigl( {\cal E}(\sigma) + {\cal E}(\tilde{\sigma}) \bigr)
\, \delta'(\sigma - \tilde{\sigma})  %%\nonumber \\
  %%&&\quad{}
- {c \over 12\pi} \delta''' (\sigma -
\tilde{\sigma})\label{eq:31origc}
\end{eqnarray}
 %%\end{mathletters}%
where we have allowed for a possible central extension of strength $c$,
and it remains to calculate this quantity. 

The gained advantage in two-dimensional space-time is that all operators
are quadratic  [see (\ref{eq:19a})--(\ref{eq:22})]; the singularity structure may be
assessed and $c$ computed; obviously it is composed of independent contributions:
\be
c = c^{\rm\,gravity} + c^{\rm\,matter} \qquad
c^{\rm\,gravity} = c_0 + c_1 \pp
\ee
Surprisingly, however, 
there is more than one way of handling infinities and more than one
answer for $c$ can be gotten.  This reflects the fact, already known to
Jordan in the 1930s \cite{8}, that an anomalous Schwinger term depends 
on how the vacuum is defined. 

In the present context, there is no argument about
$c^{\rm\,matter}$; the answer is
\be
c^{\rm\,matter} = 1 \pp  \label{eq:cmatter}
\ee
The same holds for the positively signed gravity variable
(assume $\Lambda > 0$, so that $r^1$ enters positively): 
\be
c_1 = 1 \pp \label{eq:c1}
\ee

But the negatively signed gravitational variable can be treated   in
more than one way, giving different answers for $c_0$.  The different
approaches may be named
`Schr\"odinger representation quantum field theory' 
and `BRST string/conformal field theory',
and the variety arises owing to the various ways
one can quantize a theory with a negative kinetic term, like the $r^0$
gravitational variable.  

Explicitly, the variety may be brought out by the following calculation.
Since we are dealing with quad\-ratic expressions, a direct way of
determining the Schwinger term is by normal-ordering bilinear expressions.
The field and the canonical momentum are expanded in energy eigenmodes
$e^{\pm i\omega t}$. For positively signed, conventional expressions one
calls the coefficient of the positive frequency mode ($e^{- i\omega t}$) the
annihilation operator, and `normal ordering' means put\-ting all annihilation
operators to the right of their Hermitian conjugates -- the creation
operators. This ensures that all states have positive norm and that the
spatial integral of the quadratic expression for energy is nonnegative.
Moreover, carrying out the calculation of the relevant commutator (which
requires normal reordering) exposes a {\it positive\/} Schwinger term. In
this way, for conventional, positively signed fields (matter and the `$r^1$'
gravity fields) one gets $c^{\rm matter} = c_1 = 1$. For negative-signed
fields (like the gravitational `$r^0$') it is not possible to maintain positivity
of {\it both\/} the norm and the energy. If positive norm is required -- this
is appropriate in a Schr\"odinger representation, where norms are given by
manifestly positive (functional) integrals -- one must identify the
coefficient of the {\em positive} frequency mode with the {\em creation}
operator ({\em viz.}, in a manner opposite to what was done above). This
has the consequence that the negatively signed `energy-like' expression is
negative. But in our gravitational application, this quantity does not
correspond to a physical energy, and its value is immaterial.  When this
convention is used in the normal ordering calculation of the Schwinger term,
one finds a negative quantity, $c_0=-1$. Alternatively, one may use the
coefficient of the positive frequency mode as the annihilation operator ({\em
viz.}, in a manner identical to the treatment of conventional, positively
signed expressions). Then the `energy integral' is positive, but negative
norm states are present. This is the choice taken string theory.  When the
normal ordering calculation is carried out, one finds a  positive center,
$c_0=1$, just as for positively signed quadratic expressions.  We shall make
use, in turn, of both alternatives. (The above-described variety is analogous
to what is seen in Gupta-Bleuler quantization of electrodynamics: the time
component potential $A_0$ enters with negative quadratic term.
Quantization is conventionally carried out so that negative norm states
arise, but the energy is positive; this is preferred on physical grounds.
But one could choose, alternatively, to maintain positive norm, and allow
the unphysical time-like `photons' to carry negative energy.)

In the Schr\"odinger representation quantum field theory approach, one
maintains positive norm states in a Hilbert space, and finds $c_0 = -1$,
$c^{\rm gravity} = c_0 + c_1 = 0$, $c = c^{\rm gravity} + c^{\rm matter}
= 1$.  Thus pure gravity has no obstructions, only matter provides the
obstruction.  Consequently the constraints of pure gravity can be
solved, indeed explicit formulas have been gotten by many people [9].
(This shows {\em a posteriori\/} that there can be no obstruction.)
In our formalism, according to (\ref{eq:19a})--(\ref{eq:20c}) the
constraints read
\begin{eqnarray}
{\cal E}^{\rm gravity} \ket\psi_{\rm gravity} & \sim & {\textstyle{1
\over 2}}
     \Bigl({1 \over \Lambda} {\delta^2 \over \delta r^a \delta r_a} - 
       \Lambda {r^a}' {r_a}' \Bigr)  %%\nonumber\\
 %% &&\quad{} \times 
\Psi_{\rm gravity} (r^a) = 0
\label{eq:29}\\
{\cal P}^{\rm gravity} \ket\psi_{\rm gravity} & \sim & i {r^a}'
     {\delta \over \delta r^a} \Psi_{\rm gravity} (r^a) = 0 \label{eq:30}
\end{eqnarray}
with two solutions
 %%
 %%\begin{mathletters}
\begin{equation}
  \Psi_{\rm gravity} (r^a) = {\rm exp} \pm i {\Lambda \over 2} 
       \int \dd \sigma\, \epsilon_{ab} r^a {r^b}' \pp \label{eq:31a}
  \end{equation}
This may also be presented by an action of a definite operator on the Fock
vacuum state $\ket0$,
\begin{eqnarray}
 && \Psi_{\rm gravity} (r^a) \propto %%\nonumber\\
  %% &&\qquad 
\Bigl[ \exp \pm \int
\dd  k
\,    {a_0}^{\!\dagger} (k) \, \epsilon (k) \,
    {a_1}^{\!\dagger} (-k) \Bigr] \ket0 \label{eq:31b}
\end{eqnarray}
with $  {a_a}^{\!\dagger} (k)$ creating field oscillations of definite
momentum:
\begin{eqnarray}
{a_a}^{\!\dagger} (k)  &=& {-i \over \sqrt{4 \pi \Lambda |k|}}
\int \dd \sigma \, e^{i k \sigma} \, \pi_a (\sigma) %%\nonumber\\
   %% &&\quad{} 
+ \sqrt{ {\Lambda |k| \over 4\pi} } 
\int \dd \sigma \, e^{i k \sigma} \, r^a (\sigma) \pp \label{eq:31c}
\end{eqnarray}
 %%\end{mathletters}%
As expected, the state functional is invariant against 
spatial coordinate redefinition, $\sigma \to \tilde{\sigma}(\sigma)$;
this is best seen by recognizing that the integrand in the exponent of
(\ref{eq:31a}) is a 1-form:
$d\sigma \, \epsilon_{ab} \, r^a \, {r^b}' = \epsilon_{ab} \, r^a \, dr^b$.

Although this state is here presented 
for a gravity model 
in the Schr\"odinger
representation field theory context, it is also of interest to
practitioners of conformal field theory and string theory.  The algebra
(\ref{eq:31orig})--(\ref{eq:31origc}), especially when written in decoupled
form,
\begin{equation}
\Theta_{\pm} = {\textstyle{1 \over 2}} ({\cal E} \mp P)\label{eq:decoupleA}
\end{equation}
 %%
 %%\begin{mathletters}
\begin{eqnarray}
\left[ \Theta_{\pm} (\sigma), \Theta_{\pm} (\tilde{\sigma}) \right] 
&=&   \pm i 
\left(\Theta_{\pm} (\sigma) + \Theta_{\pm} (\tilde{\sigma})\right)
  \delta' (\sigma - \tilde{\sigma})  %%\nonumber\\
  %% &&\quad{} 
\mp {ic \over 24\pi} \delta''' 
      (\sigma - \tilde{\sigma}) \label{eq:decoupleB}\\
\left[ \Theta_{\pm} (\sigma), \Theta_{\mp} (\tilde{\sigma}) \right] & = &
0 \label{eq:decoupleC}
\end{eqnarray}
 %%\end{mathletters}%
 %%
is recognized as the position-space version of the Virasoro algebra and
the Schwinger term is just the Virasoro anomaly.  Usually one does not
find a field-theoretic, nonghost realization {\it without\/} the Virasoro
center; yet the CGHS model, without matter, provides an explicit example.
Usually one does not expect that {\it all\/} the Virasoro generators
annihilate a state, but in fact our states (\ref{eq:31a})--(\ref{eq:31b}) enjoy
that property.
 %% begin insert 7a--b
Indeed, one frequently hears in string theory discussions that ``the
Virasoro anomaly is insensitive to the target space metric," {\it i.e.},
contributions from positive-signed quadratic expressions give the same
anomaly as the negative-signed ones. But our analysis shows that the
above statement is not absolute; other possibilities exist \cite{8}!

Once matter is added, a center appears, $c=1$, and the theory becomes
anomalous.  In the same Schr\"odinger representation approach used
above, one strategy is the following 
modification of a method used by Kucha\v{r} in another context
\cite{7,10}.  The Lagrange density (\ref{eq:16})
is presented in terms of decoupled constraints
(\ref{eq:decoupleA}):
 %%
 %%\begin{mathletters}
\begin{equation}
{\cal L} = \pi_a \dot{r}^a + \Pi \dot{\varphi} - \lambda^+ \Theta_+ - \lambda^-
\Theta_- \label{eq:34a}
\end{equation}
\begin{equation}
\lambda^{\pm} = \alpha \pm \beta \label{eq:34b}
\end{equation}
 %%\end{mathletters}%
 %%
Then the gravity variables $\{ \pi_a, r^a \}$ 
are transformed by a linear canonical transformation to a new set 
$\{ P_\pm, X^{\pm} \}$, in terms of which (\ref{eq:34a}) reads
 %%
 %%\begin{mathletters}
\begin{eqnarray}
{\cal L} &=& P_+ \dot{X}^+ + P_- \dot{X}^- + \Pi \dot{\varphi} %%\nonumber\\
  %%&&\quad{} 
- \lambda^+
     \Bigl( P_+ {X^+}' + \theta_+^{\rm matter} \Bigr)  %%\nonumber\\
  %% &&\qquad{} 
- \lambda^-
          \Bigl(- P_- {X^-}' + \theta_-^{\rm matter} \Bigr) \label{eq:35a}
\end{eqnarray}
\begin{equation}
\theta_{\pm}^{\rm matter} = {\textstyle\frac14} (\Pi
\pm \varphi')^2
\label{eq:35b}
\end{equation}
 %%\end{mathletters}%
 %%
The gravity portions of the constraints $\Theta_{\pm}$ have been transformed to
$\pm P_{\pm} {X^{\pm}}'$ -- 
expressions that look like momentum densities for fields
$X^{\pm}$, and thus satisfy the 
$\Theta_\pm$ algebra (\ref{eq:decoupleB})--(\ref{eq:decoupleC}) without
center, as do also momentum densities; see (\ref{eq:12a}). 

The entire obstruction in the full gravity plus matter constraints comes
from the commutator of the matter contributions $\theta_\pm^{\rm
matter}$.  In order to remove the obstruction, we modify the theory by
adding $\Delta \Theta_\pm$ to the constraint $\Theta_\pm$, such that no
center arises in the modified constraints.  An expression for
$\Delta \Theta_\pm$ that does the job is
\begin{equation}
\Delta \Theta_{\pm} = {1 \over 48 \pi} {(\ln {X^{\pm}}')}'' \pp
\end{equation}
Hence $\tilde{\Theta}_{\pm} \equiv \Theta_{\pm} + \Delta
\Theta_{\pm}$ possess no obstruction in their algebra, and can annihilate
states. Explicitly, the modified constraint equations read in the
Schr\"odinger representation (after dividing by ${X^{\pm}}'$)
\begin{eqnarray}
&&\biggl( {1 \over i} {\delta \over \delta X^{\pm}} \pm {1 \over 48
\pi{X^{\pm}}'}
        \bigl({\rm ln} {X^{\pm}}'\bigr)''  %%\nonumber\\
   %% &&\qquad\qquad\qquad 
\pm {1 \over { X^{\pm}}'} 
         \theta_{\pm}^{\rm matter} \biggr) \psi  (X^{\pm},
\varphi) = 0
\pp
\label{eq:37}
\end{eqnarray}
It is recognized that the anomaly has been removed by introducing
functional $U(1)$ connections in $X^{\pm}$ space, whose curvature
cancels the anomaly.  In the modified constraint there still is no
mixing between gravitational variables $\{ P_{\pm}, X^{\pm} \}$ and
matter variables $\{ \Pi, \varphi \}$.  But the modified gravitational
contribution is no longer quadratic -- indeed it is nonpolynomial.
Nevertheless, it is possible to analyze the consequences of (\ref{eq:37}),
by the methods that Kucha\v{r} used to solve a physically different, but
formally similar, problem \cite{10}. He presented his research at this
conference, and his analysis, which is also published\,\cite{11}, 
confirms ours\,\cite{7}.
The resulting state space describes in a diffeo\-morphism-invariant
fashion the propagation of a massless free particle on flat space. This
result follows closely what one would infer from a consideration of the
dynamics (\ref{eq:16})--(\ref{eq:18}) within classical physics. 

In the BRST quantization method, extensively employed by string and
conformal field theory investigators, one adds ghosts, which carry their
own anomaly of $c_{\rm ghost} = -26$.  Also one improves $\Theta_\pm$ by
the addition of $\Delta\Theta_\pm$ so that $c$ is increased; for
example, with
\begin{eqnarray}
\Delta \Theta_\pm &=& {Q \over \sqrt{4\pi}} \left( \Pi \pm \varphi'
\right)' \label{eq:38}
\\
c &\to& c + 3 Q^2 \pp
 \label{eq:39}
\end{eqnarray}
[The modification (\ref{eq:38}) corresponds to `improving'
the energy momentum tensor by
$(\partial_\mu \partial_\nu - g_{\mu\nu} \Box ) \varphi$.]
The `background charge' $Q$ is chosen so that the total anomaly
vanishes:
\be
c + 3Q^2 + c_{\rm ghost} = c + 3Q^2 - 26 = 0 \pp
\ee
Moreover, the constraints are relaxed by imposing that physical states
are annihilated by the `BRST' charges, rather than by the bosonic
constraints.  This is roughly equivalent to enforcing `half' the
bosonic constraints, the positive frequency portions.  In this way one
arrives at a rich and well-known spectrum.

Within BRST quantization, the negative-signed gravitational field $r^0$
is quantized so that negative norm states arise -- just as in
Gupta--Bleuler electrodynamics.   [Negative norm states cannot arise in a
Schr\"odinger representation, where the inner product is explicitly
given by a (functional) integral, leading to positive norm.]~
One then finds $c_0 = 1$; the center is insensitive to the signature
with which fields enter the action.   As a consequence, $c^{\rm gravity}
= c_0 + c_1 = 2$ so that even pure gravity constraints possess an
obstruction.

Evidently, pure gravity with $c^{\rm gravity} = 2$ requires $Q = 2
\sqrt{2}$.  The resulting BRST spectrum is 
different from
 the 
states (\ref{eq:31a})--(\ref{eq:31b}) found in the Schr\"odinger
representation. 

Gravity with matter carries $c=3$, and becomes quantizable at $Q =
\sqrt{23/3}$.  A rich spectrum emerges,  
with more states than in the Schr\"odinger approach, but with
no
apparent relation to a particle spectrum.
\bigskip
\bigskip

\section{Conclusions}\markboth{Solutions to a Quantal Gravity-Matter Field Theory on a
Line}{Conclusions}%
Without question, the CGHS model, and other similar two-dimensional
gravity models, are %%%%
  afflicted by anomalies in their constraint algebras,
which become second-class and frustrate %%%%
 straightforward quantization.
While anomalies can be calculated and are finite, their specific value
depends on the way singularities of quantum field theory are resolved,
and this leads to a variety of procedures for overcoming the problem and
to a variety of quantum field theories, with quite different properties.

Two methods were discussed: (i)~a Schr\"o\-dinger
representation with Kucha\v{r}-type
 improvement as  %%%\linebreak
 needed, {\it i.e.}, when matter is present, and (ii)~BRST quantization. 
(Actually several other approaches are also available\,\cite{12}.)  Only in
the  first method, with positive norm states and vanishing anomaly, does
the quantum theory bear a resemblance to the classical theory, in that the
classical picture of physics is reflected within the quantum theory. On the
other hand, the string theory/BRST approach yields spectra that seem quite
different from what one would expect on the basis of the classical
theory. 

Presumably, if anomalies were absent, the different quantization
procedures (Schr\"odinger representation, BRST, $\ldots$) would
produce the same physics.  However, the anomalies {\it are\/} present and
interfere with equivalence.

Finally, I remark that our investigation has exposed an interesting
structure within Virasoro theory: there exists a field-theoretic
realization of the algebra without the anomaly, in terms of spinless
fields and with no ghost fields.   Moreover, there are states that are
annihilated by {\it all\/} the Virasoro generators.

What does any of this teach us about the physical four-dimensional
model?  I believe that an extension in the constraint algebra will
arise for all physical, propagating degrees of freedom: for matter
fields, as is seen already in two dimensions, and also for gravity
fields, which in four dimensions (unlike in two) carry physical energy.
How to overcome this higher-dimensional obstruction to quantization is
unclear. 
Especially problematic is the fact that flat-space calculations of anomalous
Schwinger terms in four dimensions yield infinite results, essentially for
dimensional reasons. Therefore,   any 
announced `solutions' to the constraints that result from
{\it formal\/} analysis must be viewed as preliminary:
properties of the Hilbert space and of the inner
product must be fixed first in order to give an unambiguous
determination of any obstructions.

I believe that our two-dimensional investigation, although in a much simpler
and unphysical setting, nevertheless contains important clues for
realistic theories.  Certainly that was the lesson of vector gauge theories:
anomalies and vacuum angle have corresponding roles in the Schwinger
model and in QCD!


\newpage
\begin{thebibliography}{99}

\frenchspacing

\bibitem{1}
See R.~Jackiw, 
 in
{\it Quantum Mechanics of Fundamental Systems},
C.~Teitelboim, ed.
(Plenum, New York NY, 1988).

\bibitem{2}
D.~Boulware and S.~Deser,
{\it J.~Math. Phys} {\bf 8}, 1468 (1967).

\bibitem{3}
R.~Jackiw and C.~Teitelboim, in
{\it Quantum Theory of Gravity},
S.~Christensen, ed.
(A.~Hilger, Bristol UK, 1984).

\bibitem{4}
D.~Cangemi,
{\it Phys. Lett.} {\bf B297}, 261 (1992);
A.~Ach\'ucarro, 
{\it Phys. Rev. Lett.} {\bf 70}, 1037 (1993).

\bibitem{5}
A.~Polyakov, 
{\it Mod. Phys. Lett.} {\bf A2}, 893 (1987),
{\it Gauge Fields and Strings}
(Harwood, New York NY, 1987).

\bibitem{6}
C.~Callan, S.~Giddings, J.~Harvey, and A.~Strominger,
{\it Phys. Rev. D} {\bf 45} 1005 (1992);
H.~Verlinde in
{\it Sixth Marcel Grossmann Meeting on General Relativity},
M.~Sato and T.~Nakamura, eds.
(World Scientific, Singapore, 1992).

\bibitem{7}
Our two most recent papers on this entire subject, with reference to earlier
work, are  
D. Cangemi,
R.~Jackiw, and B.~Zwiebach,  {\it Ann. Phys.} (NY) {\bf 245}, 408 (1996);
 E. Benedict, R. Jackiw, and H.-J. Lee,  
{\it Phys. Rev. D} {\bf 54}, 6213 (1996).

\bibitem{8}
P.~Jordan, {\it Z. Phys.} {\bf 93}, 464 (1935).
In the context of the energy-momentum tensor commutator anomaly, this is
explained in R.~Floreanini and R.~Jackiw,
{\it Phys. Lett.} {\bf B175}, 428 (1986).

\bibitem{9}
D.~Cangemi and R.~Jackiw,
{\it Phys. Lett.} {\bf B337}, 271 (1994),
{\it Phys. Rev. D} {\bf 50}, 3913 (1994);
D.~Amati, S.~Elitzur, and E.~Rabinovici,
{\it Nucl. Phys.} {\bf B418}, 45 (1994);
D.~Louis-Martinez, J.~Gegenberg, and G.~Kunstatter, 
{\it Phys. Lett.} {\bf B321}, 193 (1994); 
E.~Benedict, 
{\it Phys. Lett.} {\bf B340}, 43 (1994);
T.~Strobl, 
{\it Phys. Rev. D} {\bf 50}, 7346 (1994); Ref.\,\cite{7}.

\bibitem{10}
K.~Kucha\v{r}, {\it Phys. Rev. D} {\bf 39}, 2263 (1989);
K.~Kucha\v{r} and G.~Torre, {\it J. Math. Phys.} {\bf 30}, 1769 (1989). 

\bibitem{11}
K.~Kucha\v{r}, J. Romano, and M. Varadarjan,
 {\it Phys. Rev. D} {\bf 55},
 xxxx (1997).

\bibitem{12}
Cangemi, Jackiw, and Zwiebach, Ref.\,\cite{7}

\end{thebibliography}
\end{document}